\documentclass{emulateapj}

\begin{document}
\title{A Variable Black Hole X-Ray Source in a NGC~1399 Globular Cluster\altaffilmark{1}}

\author{I Chun Shih\altaffilmark{2,3}, Arunav Kundu\altaffilmark{2}, Thomas J. Maccarone\altaffilmark{4}, Stephen E. Zepf\altaffilmark{2}, \& Tana D. Joseph\altaffilmark{4}}

\altaffiltext{1} {Based on observations made with the Chandra X-ray Observatory, which is operated by the Smithsonian Astrophysical Observatory for and on behalf of the National Aeronautics Space Administration under contract NAS8-03060
and the NASA/ESA Hubble Space Telescope which is operated by
    the Association of Universities for Research in Astronomy, Inc., under NASA contract
    NAS 5-26555.}

\altaffiltext{2}{ Physics \& Astronomy Department, Michigan State University, East Lansing, MI 48824; e-mail: Stephen.Shih@obspm.fr, akundu@pa.msu.edu, zepf@pa.msu.edu}
\altaffiltext{3}{Current address: Institute of Astronomy, National Tsing Hua University, Hsinchu, Taiwan}
\altaffiltext{4}{School of Physics \& Astronomy, University of Southampton, Southampton, Hampshire SO17 1BJ; e-mail: tjm@astro.soton.ac.uk, tdj1f08@soton.ac.uk }

\begin{abstract}

We have discovered an accreting black hole (BH) in a spectroscopically confirmed globular cluster (GC) in NGC 1399 through monitoring of its X-ray activity. The source, with a peak luminosity of L$_{X}$$\simeq$2$\times$10$^{39}$ergs~s$^{-1}$,  reveals an order of magnitude change in the count rate within $\simeq$10 ks in a Chandra observation. The BH resides in a metal-rich [Fe/H]$\simeq$0.2 globular cluster. After RZ~2109 in NGC 4472 this is only the second black-hole X-ray source in a GC confirmed via rapid X-ray variability.  Unlike RZ~2109, the X-ray spectrum of this BH source did not change during the period of rapid variability. In addition to the 
short-term variability the source also exhibits long-term variability. After being bright for at least a decade since 1993 within a span of 2 years it became progressively fainter, and eventually undetectable, or marginally detectable, in deep Chandra and XMM observations. The source also became harder as it faded. The characteristics of the long term variability in itself provide sufficient evidence to identify the source as a BH. 
The long term decline in the luminosity of this object was likely not recognized in previous studies because the rapid variability within the bright epoch suppressed the average luminosity in that integration. The hardening of the spectrum accompanying the fading would also make this black hole source indistinguishable from an accreting neutron star in some epochs. Therefore some low mass X-ray binaries identified as NS accretors in snapshot studies of nearby galaxies may also be BHs. Thus the discovery of the second confirmed BH in an extragalactic GC through rapid variability at the very least suggests that accreting BHs in GCs are not exceedingly rare occurences.

\end{abstract}


\section{Introduction}
The high spatial resolution of X-ray telescopes such as Chandra and XMM-Newton has led to the discovery of large populations of resolved X-ray sources around nearby galaxies. The vast majority of the point sources observed in elliptical galaxies are believed to be low-mass X-ray binaries (LMXBs). Recent studies have established that roughly 40\% of these LMXBs are associated with globular clusters \citep[e.g.][]{kundu07,sivakoff07}. This overabundance of LMXBs in globular clusters (GCs), that typically account for less than $\sim$0.1\% of the stellar mass of a galaxy, reflects the importance of stellar dynamical interactions in the dense inner regions of clusters that leads to the enhanced formation of close binaries.

This raises the interesting question of whether some of these LMXBs may be black hole (BH) accretors. There is much interest in black holes residing in globular clusters in part because it is often suggested that GCs are among the most likely hosts of intermediate mass black holes. On the other hand there have also been suggestions that most black holes are ejected from GCs due to dynamical processes \citep{sigurdsson93,kulkarni93}. 

One way to detect a black hole in a GC is to search for a characteristic X-ray signature if the BH happens to be accreting from a nearby companion. So far no such accreting BHs have been observed in any of the globular clusters in the Milky Way. But  Chandra and XMM observations have identified a number of GCs in nearby galaxies that host LMXBs that are brighter than the Eddington luminosity for an accreting neutron star \citep{angelini01,kundu02,kim06}. However these bright sources may also represent the superposition of many neutron star LMXBs \citep[e.g.][]{angelini01,kundu07} in globular clusters. Detection of rapid X-ray variability in such sources may be the only definitive way to confirm a black hole accretor with present instruments \citep[e.g.][]{kalogera04}.

We have previously discovered the first, and so far only, black hole accretor in a GC through X-ray variability in RZ~2109, a globular cluster in the bright Virgo cluster galaxy NGC 4472 \citep{maccarone07,shih08,zepf08}. Identification of a single such candidate leaves open many intriguing questions about the ubiquity of such systems. It is  possible that the discovery of a BH in the globular cluster RZ~2109 owes to an extraordinarily fortuitous set of circumstances. In this paper we demonstrate the presence of a black hole in a second extragalactic globular cluster in NGC 1399 through variability in the X-ray flux of a luminous accretor.

NGC~1399 is the giant elliptical galaxy at the center of the Fornax Cluster. It has a rich and well-studied globular cluster system \citep[e.g.][]{grillmair,dirsch03} with a bimodal GC metallicity distribution that is typical in such galaxies \citep[e.g.][]{kundu01}. Many joint optical and X-ray studies of NGC 1399 have showed that it has one of the highest fraction LMXBs associated with GCs of any galaxy studied to date. As in other galaxies metal-rich clusters are more likely to host LMXBs than metal-poor ones \citep{angelini01,kim06,kundu07}. We describe below the discovery of a black hole in one of the luminous GCs in this galaxy.

\section{Observations and Results}

\subsection{X-rays}

NGC~1399 has been observed multiple times by both the Chandra and XMM-Newton observatories in the past decade (Table 1). This provides an unique opportunity to trace X-ray activity in LMXBs, and in particular those associated with GCs for the purpose of this study.

 We downloaded the archival Chandra and XMM-Newton data  of NGC 1399 listed in Table 1 and processed it following standard data reduction guidelines using {\tt CIAO v4.0} and {\tt SAS v8.0}, respectively. Intervals of high background were filtered out from the data, and all images were corrected for exposure and vignetting. We adopted the wavelet method of source detection to identify point sources. 

 We discovered that one of the candidate LMXBs observed during the course of Chandra observation 319  displayed  rapid X-ray variability on the timescale of $\sim$10 ksec (Fig 1). A literature search reveals that this source has been observed by ROSAT-HRI multiple times between 1993 and 1997 and is also listed in the 2XMM catalogue as 2XMM J033831.7-353058:042943 \citep{Paolillo02, liu05, watson09}. This object was also identified  in \citet{kundu07} as X-ray source number 157, CXOKMZJ033831.7-353058 (hereafter, the source). The 5.4$\times$10$^{38}$ ergs s$^{-1}$ X-ray luminosity derived using the average count rate of the source through the entire Chandra observation is brighter than the Eddington luminosity of an accreting neutron star (also see \S2.4.1 below), suggesting that this may be a black hole candidate. However the source lies outside the HST image of the inner region analyzed in \citet{kundu07}. Thus, we turn to other data sets to check whether this object is associated with a globular cluster.

\subsection{Optical Counterpart}

We downloaded F606W band HST-ACS images of NGC 1399 from the archive (HST GO proposal 10129) and created an aligned mosaic of the nine independent pointings of the region around the galaxy. We used the Multidrizzle routine based on the drizzle code of \citet{fruchter02} for this purpose, taking into account the small errors in the default header values. We aligned the HST mosaic to the USNO-A2.0 astrometric system \citep{monet98}. Next we identified the point sources corresponding to the spectroscopically confirmed GCs studied by \citet{dirsch04}. The location of the GCs in the HST mosaic image was used to derive more secure positions for these objects. Finally we spatially aligned the X-ray and optical images in the manner described in \citet{maccarone03}. The rms of the optical and X-ray source matches was 0.2$\arcsec$.

	The variable X-ray source CXOKMZJ033831.7-353058 is coincident with the 
globular cluster identified as 86:53.0 by \citet{dirsch04}. The matching radius is only 0.14$\arcsec$ confirming that this object is located in a GC. It is located almost exactly 4' due South of NGC 1399 at RA=03:38:31.7 Dec=-35:30:59.21 (J2000). These coordinates are based on the location of the source in the HST image aligned to the USNO system. The host globular cluster has a magnitude and color R=22.02$\pm$0.03 and C-R=2.04$\pm$0.09 \citep{dirsch04}. In the next few sections we examine all the Chandra and XMM-Newton observations of NGC~1399 which covered the position of the source (Table 1). 

\subsection{Short Term Variability and Spectral Analysis}

\begin{figure}
\begin{center}
\includegraphics[width=90mm]{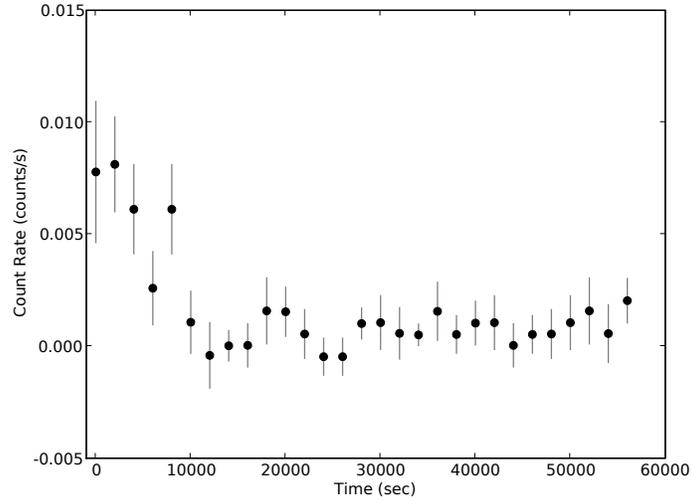}
\caption{The Chandra ACIS-S3 binned lightcurve of the globular cluster X-ray source in NGC 1399 from the 2000-1-18 Obs ID 319 obervations. Photons in the energy 0.3 to 8.0 keV are incuded, and the count rate is averaged over 2,000s.}
\label{lc}
\end{center}
\end{figure}

We created background-subtracted lightcurves from the observations in which the source can be detected. As the source is fairly isolated with no obvious X-ray emitters nearby we used an annular region surrounding the object with $\sim$1:1 area ratio to determine the background-subtracted count rate. Only the first data set,  Chandra observation 319, revealed a rapid short-term variability (see Fig. \ref{lc}). The count rate declined  by an order of magnitude during the first 10 ksecs. The source remained faint for the rest of the observation and showed no further evidence of measurable variability. To verify that the decline was not due to an instrumental effect we compared the lightcurves of other X-ray sources in the same CCD (ACIS-S3), and found no similar patterns. In the rest of this paper we consider the first 9,873 s of Chandra Obs ID 319 in which the count rate of the is either high or declining the ``bright" phase. The other 46,070 s during which the count rate is at a steady low level is referred to as the ``faint" phase.

Spectroscopic analysis of the Chandra data was carried out using the Interactive Spectral Interpretation 
System (ISIS), version 1.4.9-55.  Data 
were binned within ISIS into groups with signal-to-noise of at least 2.0, 
and all channels with at least this signal-to-noise within the range from 
0.3-8.0 keV were included in the fits. Due to the limited count rate we only attempted to fit two standard models which are commonly used to describe the continuum spectra of X-ray binaries, disk blackbody  ({\tt DISKBB}) and power-law ({\tt PO}). The foreground absorption was fixed to the Galactic neutral Hydrogen column density (nH) in the direction of NGC~1399, $\sim$ 1.31$\times$10$^{20}$ cm$^{-2}$ \citep{dickey90}.

 The data were fit using the 
Gehrels statistic to define the errors (with the errors set to 
1+$\sqrt{(N+.75)}$, where $N$ is the number of counts in the bin) on the 
individual bins, as this binning has been shown to provide more reliable 
fitting to bins with low count numbers.  It was found that binning to the 
more standard signal-to-noise of 5.0 per bin led to too few channels for 
fitting and binned out real information in the data. The uncertainty estimates for all fits reflect 90\% confidence limits.

The results of the spectral fitting of the bright and faint epochs of the Chandra 319 observations are summarized in Table 2 and plotted in Fig 2.  Both the disk blackbody and power law fits reveal a soft spectrum for both time periods, with the PO model providing a marginally better fit. This is consistent with a luminous X-ray binary in a very-high state, where the power law component dominates the luminosity  \citep{nowak95,mcclintock06}. We note here that some of the inordinately large uncertainty estimates in the normalization of the disk blackbody models, for these and other fits listed in Table 2, are correlated to the large errors in the temperature, but lead to consistent total luminosity.

  There is no statistically significant difference between the spectral properties of the source in the two epochs.  Using the power-law model the X-ray flux of the source in the energy range of 0.3-8.0 keV is estimated to be 2.8$\times$10$^{-14}$ ergs\ cm$^{-2}$\ s$^{-1}$ in the bright phase and 2.8$\times$10$^{-15}$ in the faint phase. Adopting a distance of 20 Mpc \citep{tonry01} this implies an isotropic luminosity of 1.3$\times$10$^{39}$ ergs\ s$^{-1}$ during the bright epoch and 1.3$\times$10$^{38}$ ergs\ s$^{-1}$ during the faint one. We note here that the average luminosity quoted in \citet{kundu07}, and mentioned in \S 2.1 above was calculated using fixed a spectral index $\Gamma$=1.7 for all sources in NGC 1399 due to the difficulty of reliably measuring this parameter for many of the sources superimposed on the bright hot gas in the inner regions.

\subsection{Long Term Variability}

We searched for the source in all of the Chandra and XMM data sets listed in Table 1. The source was detected in all observations through the middle of 2003 but appears to have subsequently turned off indicating long term variability. The individual observations are briefly discussed below.

\subsubsection{XMM Observation 0055140101}

These XMM observations were taken one year after the Chandra 319 observations discussed above. The source is located on the edge of the XMM-Newton EPIC MOS2 image and is the basis for its detection and listing as 2XMM J033831.7-353058:042943 in the second XMM-Newton Serendipitous Source Catalogue \citep[2XMM,][]{watson09}. Following standard data reduction guidelines using  {\tt SAS v8.0}. As in the Chandra analysis we used an annular region surrounding the object with $\sim$1:1 area ratio to account for the background.

We fit the XMM  data of the source in XSPEC 11.0 using the Gehrels statistic, without binning channels. Within the uncertainties the spectral properties of the source are indistinguishable from the Chandra 319 observations (Table 2). The flux, $F_{X}\simeq$4.2$\times$10$^{-14}$\ ergs\ cm$^{-2}$\ s$^{-1}$ for the power law fit, is similar to the level of the bright phase of the Chandra 319 observations.  Thus the source regained its luminosity within a year of the short term variability observed in the earlier Chandra data. This translates to an isotropic luminosity of L$_X$$\simeq$2$\times$10$^{39}$\ ergs\ s$^{-1}$, which is slightly brighter than the estimate from the bright phase of Chandra Obs ID 319, but consistent with previous ROSAT estimates \citep{liu05}. However, we note that the short timescale variation in Chandra Obs ID 319 occurred at the very beginning of the observation leaving open the possibility that the source was in fact brighter immediately prior to this observation. Thus we consider the L$_X$$\simeq$2$\times$10$^{39}$\ ergs\ s$^{-1}$ luminosity from this XMM observation to be a fairer estimate of the peak brightness of this source. 

\subsubsection{XMM Observation 0012830101}

The source was detected in the 10 ks XMM Newton observations of NGC 1399 on 2001-06-27 (Obs ID 0012830101). Given the short exposure and consequently low counts we do not attempt to fit the spectrum. The 2XMM flux of this source is estimated to be 
4.1$\times$10$^{-14}$\ ergs\ cm$^{-2}$\ s$^{-1}$, which implies an isotropic luminosity of L$_X$$\simeq$1.9$\times$10$^{39}$\ ergs\ s$^{-1}$. Thus the source appears to have remained as bright as observed in the previous XMM observation obtained 6 months before.

\subsubsection{Chandra Observation 2942}

The source was in the field of view of the Chandra ACIS observation of NGC 1404, taken roughly three years after the  Chandra 319 observation. At this time, the spectrum was  harder than at earlier epochs (Table 2). The X-ray flux in the 0.3-8.0 keV band from the power law fit is 1.9$\times$10$^{-14}$\ erg\ cm$^{-2}$\ s$^{-1}$, a factor of $\sim$2 lower than the previous epoch. However, the harder spectrum and the kT$\simeq$0.9 keV estimate suggests that the source may have undergone a state transition to a high/soft state as observed in stellar mass black holes \citep{nowak95,mcclintock06} and may be better described by a disk blackbody model. The flux estimate of $F_{X}\simeq$1$\times$10$^{-14}$\ erg\ cm$^{-2}$\ s$^{-1}$ from a disk blackbody fit implies a luminosity of 4.8$\times$10$^{38}$\ ergs\ s$^{-1}$, and a factor of 4 decline in the brightness from the previous XMM epoch.

\subsubsection{Chandra Observations 4172 \& 4174}

Both of these observations were part of the Fornax Cluster survey. The source was in the field of view of the Chandra ACIS I3 (4172) or I0 (4174) chips. Although it was detected by CIAO's source detection tool, the count rate in each case was too low (3.06$\times$10$^{-4}$ and 1.39$\times$10$^{-4}$ cts/s) to derive meaningful result by fitting the spectrum. Instead we determined the X-ray flux using the simulation tool {\tt PIMMS} by assuming that the spectral properties were unchanged since the Chandra observation 2942. The X-ray flux in the 0.3-8.0 keV range is $\simeq$3.7$\times$10$^{-15}$\ erg\ cm$^{-2}$\ s$^{-1}$. Thus the luminosity of the source has dropped by another factor of 2 to 1.8$\times$10$^{38}$\ ergs\ s$^{-1}$ luminosity since the Chandra 2942 observation.
  
\subsubsection{XMM Observations 0304940101 \& 0400620101}
The source is not detected in the XMM-Newton observations 0304940101 and 0400620101. Neither automated routines nor a careful visual inspection revealed any evidence of an object at or near the location of the source. The measured count rate at the position of the source is indistinguishable from that of the nearby background. Using the XMM FLIX tool we estimate an upper limit of 
2.7$\times$10$^{-14}$\ ergs\ cm$^{-2}$\ s$^{-1}$ on July 30, 2005 and 1$\times$10$^{-14}$\ ergs\ cm$^{-2}$\ s$^{-1}$ on Aug 23, 2006. Thus the source was fainter than $L_{X}\leq$4.8$\times$10$^{38}$\ ergs\ s$^{-1}$ at this time.

\subsubsection{Chandra Observations 9798 \& 9799}

We also investigate the two Chandra observations 9798 and 9799 aimed at SN2007on.  Since they were
taken only three days apart, it makes sense to combine the results from the analyses of the two data sets to get one flux value/upper limit. The black hole candidate is 4.9' off axis in these observations, but is undetected in both. We attempt to estimate an upper limit on the flux by
determining how many photons are within 2$\arcsec$ of its location. In observation
9798 we find 4 photons from 0.5 to
8.0 keV within that radius, with an expectation value of $0.96\pm.03$
photons estimated by looking at an off-axis background region.  In
observation 9799 we find 7 photons,
with an expectation value of $1.24\pm0.03$.  The source is thus marginally
detected when the two observations together are considered -- there is only
a $2\times10^{-5}$ probability that 11 photons will be detected given a
background expectation value of 2.2 photons. However, if one considers the trials
incurred by checking all positions on the ACIS-S3 chip the
detection is not statistically significant.

One can also make an estimate of the upper limit to the luminosity of the
source during these two observations, given that the source is marginally
detected, with 11 photons seen in 40 ksecs.  The 99\% confidence level
upper limit on the underlying count rate is then 20.5 photons, of which 18.3
would be expected to be source photons.  This translates to a count rate of
$4.6\times10^{-4}$ photons sec$^{-1}$. Assuming the spectrum was unchanged since the Chandra 2942 observations this corresponds to a flux limit of $4.6\times10^{-15}$  ergs\ cm$^{-2}$\ s$^{-1}$ and a  luminosity upper limit $L_{X}\leq$$2.2\times10^{38}$ ergs s$^{-1}$.

\subsubsection{Chandra Observation 9530}

The source was in the field of view of the Chandra 9530 observation of NGC 1399 obtained on 2008-06-08. No object is detected at the location of the source with the probability of false detection set to 10$^{-6}$. However there is a marginal detection  when the false source probability is relaxed to 10$^{-4}$. This yields 13 background subtracted counts during the course of this 60 ks observation. Although is insufficient to fit a spectrum, we can estimate the luminosity by comparing the count rate to previous Chandra observations. If the source had a soft spectrum similar to that observed on 2001-1-18 then scaling the count rate to the Chandra 319 observations yields a luminosity of $L_{X}\leq$$4.1\times10^{37}$ ergs s$^{-1}$. If on the other hand it retained the harder spectral shape of the 2003-2-13 Chandra 2942 detection, scaling the count rate to that observation estimates a luminosity of $L_{X}\leq$$4.8\times10^{37}$ ergs s$^{-1}$. Thus, irrespective of the detils of the spectrum the source was $L_{X}\leq$$4.8\times10^{37}$ ergs s$^{-1}$ or fainter at this epoch. This is fainter than the upper limit of the luminosities derived from the previous observations in which the source was not detected, and confirms that it faded since the 2003 Chandra detection. we also note that a luminosity of $L_{X}\leq$$4.8\times10^{37}$ ergs s$^{-1}$ is typical of bright neutron star accretors. There is a high probability of such an object being associated with a high mass, metal-rich globular cluster such as the one being studied here \citep{kundu07}. Thus the black hole accretor itself may in face be fainter than the luminosity estimated from the Chandra 9530 observation.

\subsection{Discussion and Analysis}

We have confirmed that the X-ray source previously identified as CXOKMZJ033831.7-353058 \citep{kundu07} and 2XMM J033831.7-353058:042943 \citep{watson09} is associated with the spectroscopically  confirmed globular cluster 86:53.0 \citep{dirsch04} in NGC 1399. We have examined the X-ray activity of the source over a period of 8 years using archival Chandra and XMM-Newton data.

Between Jan 2000 and Feb 2003 the X-ray luminosity of the source was above 10$^{39}$\ ergs\ s$^{-1}$, based on an estimated distance of 20 Mpc to NGC 1399 \citep{tonry01}. Earlier ROSAT observations sampling various epochs between 1993 and 1997 estimated a similar luminosity of 2$\times$10$^{39}$\ ergs\ s$^{-1}$ \citep{liu05}. This is well beyond the Eddington luminosity of spherical accretion on to a neutron star and places the source into the category of an ultra luminous X-ray source.\\

During the course of the first Chandra observation in this analysis (Obs ID 319 in Jan 2000) the source revealed a rapid short-term decline in the count rate within 10 ksecs, demonstrating that the source is not a superposition of multiple neutron star binaries and confirming that it is a black hole. After RZ~2109 in NGC 4472  \citep{maccarone07} this is only the second BH X-ray accretor positively identified through short term variations in the count rate. Curiously both sources revealed roughly order of magnitude declines over similar time periods of $\simeq$10 ks. 

But the parallels appear to end there. The decline in the count rate in RZ~2109 was primarily in the soft X-ray band, indicating a change in the X-ray spectrum consistent with foreground absorption and not a variation in the inherent X-ray luminosity of the source \citep{shih08}. On the other hand there is no statistical difference in the spectrum of CXOKMZJ033831.7-353058 between the bright and faint phases in the Chandra 319 observations (Table 2 \& Fig. 2). This suggests that either the luminosity of the source changed during the course of the Chandra319 observation or that a portion of the X-ray emitting region was obscured by an eclipsing event. 

The NGC 1399 BH also reveals long term variability. Other than a portion of the Chandra Obs ID 319 integration the sources remained brighter than 10$^{39}$\ ergs\ s$^{-1}$ from the 1993 ROSAT observations through the 2001 XMM-Newton observations. Thereafter it became progressively fainter. It was not detected in XMM and Chandra observations between 2005 and 2007. A $L_{X}\leq$$4.8\times10^{37}$ ergs s$^{-1}$ source was marginally detected in a deeper 2008-06-08 Chandra observation. However such a luminosity is typical of bright neutron star accretors ad there is a high probability of such an object being associated with the host globular cluster of this source. The black hole accretor itself may in fact be even fainter than $L_{X}=$$4.8\times10^{37}$ ergs s$^{-1}$ at this point.

 The Chandra 2942 observation from 2003 which showed the first clear evidence of a long term decline in the luminosity has a spectral index of $\Gamma$=1.3, considerably harder than the $\Gamma$=2.4-2.9 spectrum at earlier epochs. This suggests that the source underwent a state transition from a very-high state where the power law component dominates to a high/soft state where the emission is primarily thermal   \citep{nowak95,mcclintock06}. Such state changes accompanying luminosity variations in a GC BH points to some interesting issues about identifying and quantifying BHs in globular clusters.

	BH candidates in nearby galaxies (and their GC systems) are identified on the basis of their higher luminosities and softer spectra as compared to NS accretors. However, the hardening of the spectrum of CXOKMZJ033831.7-353058 as it faded places both its luminosity and its spectrum in the range of typical neutron star LMXBs (L$_{X}$$\leq$10$^{38}$ergs~s$^{-1}$, $\Gamma$$\simeq$1.7) in the latter epochs of our survey. Thus an unrecognized fraction of the LMXBs in GCs tagged as neutron star systems in snapshot observations of extragalactic systems are likely BH accretors.

 The long term variability of CXOKMZJ033831.7-353058 itself provides strong evidence that the source is a black hole. The source was consistently in the ultraluminous regime for at least a decade prior to 2003, before fading relatively rapidly within a span of less than two years. If this object were a superposition of multiple close to Eddington limited neutron star LMXBs it is highly unlikely that they would all conspire to change luminosity within such a short time period. Moreover, given the fact that this is similar to the light travel time across the core of a typical cluster (where neutron star LMXBs are expected to reside) the degree of coincidence required is even more unlikely. The fact that a confirmed X-ray bright black hole system in a GC undergoes physical changes similar to those seen in stellar mass BHs in the MW on the timescales of years is encouraging in terms of identifying and studying the demographics BHs in GCs. It indicates that BHs can be securely identified by monitoring on month-to-year scales without requiring fortuitous occurances of short term variations within a single observation.

There have been previous attempts to detect BHs in elliptical galaxies by such monitoring  \citep[e.g.][]{irwin06,brassington08}. However, the only likely BHs identified using this technique \citep{fabbiano06,sivakoff08} have been field sources that are not associated with GCs. It is particularly interesting that \citet{irwin06} which studied NGC 1399 using a subset of this data set did not recognize the transient nature of CXOKMZJ033831.7-353058. The luminosity of the source varied by about an order of magnitude between the (bright phase of the)  Chandra 319 observations and Chandra 4172 observations that bookend the \citet{irwin06} analysis. This can likely be attributed averaging across the short term variation in the first epoch which leads to an underestimate of the consistently higher luminosities observed over a decade. We note that the luminosity quoted in our own \citet{kundu07} study similarly underestimated the brightness of CXOKMZJ033831.7-353058 in the Chandra 319 integration for the same reason. Given that at least four accreting BH systems in elliptical galaxies \citep[][and this source]{fabbiano06,maccarone07,brassington10} are known to show short term X-ray flux variability (accompanied by spectral changes in some cases) this strongly argues that careful analysis of both short and long term variations of potential BH candidates must be carried out in concert if more BHs are
to be identified by this method.

Another feature of accreting BHs in globular clusters may be the presence of optical emission lines. The optical spectrum of RZ~2109 reveals broad, $\simeq$2000 km s$^{-1}$ wide, bright [O III] emission associated with the BH \citep{zepf08}. \citet{irwin09} argue that another ultraluminous X-ray source in a globular cluster in the inner region of NGC 1399 is a black hole on the basis of $\simeq$75 km s$^{-1}$ [O III] and [N II] lines. An optical spectrum of the host GC of our source taken with the SOAR optical telescope in 2009 shows no evidence of emission lines. A previous 2-hour Magellan/IMACS optical spectrum of this globular cluster taken in November 2006 (Jimmy Irwin, private communication) also reveals no emission lines. 
However we note, on the basis of our X-ray monitoring, that the X-ray source had faded considerably before either set of optical spectra were obtained. 

The host GC of CXOKMZJ033831.7-353058 also differs from RZ~2109 in NGC 4472 in important ways. While RZ~2109 is a metal-poor cluster, 86:53.0 in NGC 1399 \citep{dirsch04} has a color of C-R=2.04$\pm$0.09. 
This places it at the reddest, and hence the most metal-rich end of the metal-rich subpopulation of GCs \citep{dirsch04}. 
Based on the  calibration of \citet{lee08} this translates to a supersolar metallicity of [Fe/H]$\simeq$0.2 dex.
  RZ~2109 is also one of the most luminous, and hence most massive GCs in NGC 4472. While 86:53.0 is also among the largest GCs in NGC 1399, unlike RZ~2109 it is not at the tail end of the luminosity function. In quantitative terms RZ~2109 is about 2.6 magnitudes brighter than the peak of the globular cluster luminosity function while 86:53.0 in NGC 1399 is only 1.3 magnitudes brighter than the peak. In other words 86:53.0 is about one-third the 2$\times$10$^6$M$_\odot$ mass of RZ~2109 \citep{zepf08}. 

Our discovery of the second confirmed BH accretor in a globular cluster on the basis of X-ray variability indicates that the discovery of RZ~2109 in NGC 4472 was not due to a particularly fortuitous configuration in that system and provides encouraging evidence that more such BHs can be identified by monitoring the X-ray activity. Further studies of temporal variability of luminous X-ray sources  in nearby galaxies are needed to establish reliable statistics on the ubiquity of black holes in globular clusters.

\acknowledgments{ICS, SEZ and AK acknowledge support from NASA through ADP grant NNX08AJ60G and Chandra grant AR8-9011X. AK was also supported by NASA LTSA grant NAG 5-12975. T.J. acknowledges support from a Stobie-SALT studentship, funded jointly by the NRF of South Africa, the British Council and Southampton University. We thank the referee Jimmy Irwin for providing helpful comments that improved this paper.}

\clearpage
\phantom{a} 

\begin{deluxetable}{llllll}
\tablewidth{0pt}
\tablecaption{Chandra and XMM-Newton Observations of the Source}
\startdata
\tableline
\tableline
  Date & Observatory & Obs ID & Instrument & Effective Exposure Time & Detection \\ 
\tableline
  2000-01-18 & Chandra 	& 319 		& ACIS-S3 	& 55,943s & yes \\
  	&	&	&		& (bright:9,873s; faint:46,070s) & \\
  2001-01-07 & XMM 	& 0055140101$^1$ & EPIC-MOS2 	& 47,080s & yes\\
  2001-06-27 & XMM 	& 0012830101 	& EPIC-MOS2, \& PN & 10,085s & yes\\
  2003-02-13 & Chandra 	& 2942 & ACIS-S3 & 29,420s 			& yes\\
  2003-05-26 & Chandra 	& 4172 & ACIS-I3 & 44,503s 			& yes\\
  2003-05-28 & Chandra 	& 4174 & ACIS-I0 & 45,672s 			& yes\\
  2005-07-30 & XMM 	& 0304940101 	& EPIC-MOS2, PN & 20,007s 	& no\\
  2006-08-23 & XMM 	& 0400620101 	& EPIC {\it all}$^{2}$ & 73,040s & no\\
  2007-12-24 & Chandra 	& 9798 		& ACIS-S3 	& 18,302s 	& no\\
  2007-12-27 & Chandra 	& 9799 		& ACIS-S3 	& 21,288s 	& no\\
  2008-06-08 & Chandra 	& 9530 		& ACIS-S3 	& 59,348s 	& yes$^{3}$\\
\enddata
\tablenotetext{1}{2XMM catalogue \citep{watson09}.}
\tablenotetext{2}{MOS1, MOS2 \& PN cameras.}
\tablenotetext{3}{The luminosity of this marginal detection is lower than the upper limits implied in the previous non-detections, and is in the range of typical neutron star accretors. See text for details.}

\end{deluxetable}

\begin{deluxetable}{lllllll}
\tighten
\tablecaption{Spectral Fitting Results and Flux Estimates}
\label{spectral fitting}
\startdata
\tableline
\tableline
                \multicolumn{6}{c}{\underline{Disk Blackbody}}              \\
Date       & Obs. ID/Epoch          	& kT$_{in}$  		& Norm. 			          & $\chi_{\nu}^{2}$ & Unabsorbed Flux \\
       &          	& (keV) 		& 			          &  & ergs cm$^{-2}$ s$^{-1}$ \\
\tableline
2000-1-18  & Chandra 319 (bright) 	& 0.29$\pm^{0.2}_{0.09}$ 	& 0.17$\pm^{2}_{0.2}$ 	             & 0.44  & 1.9$\times$10$^{-14}$ \\  
2000-1-18  & Chandra 319 (faint) 	& 0.21$\pm^{0.3}_{0.1}$ 	& 0.077$\pm^{7}_{0.08}$                & 0.62 & 2.1$\times$10$^{-15}$ \\
2001-1-07  & XMM 0055140101 	        & 0.42$\pm^{0.3}_{0.2}$ 	& 0.061$\pm^{0.6}_{0.06}$               & 0.26  & 3.3$\times$10$^{-14}$ \\
2003-2-13  & Chandra 2942 		& 0.90$\pm^{3}_{0.5}$ 	& 7.74$\pm^{19000}_{8}$$\times$10$^{-4}$ & 0.66& 1.0$\times$10$^{-14}$ \\
\tableline
\tableline
        	\multicolumn{6}{c}{\underline{Power Law}} 	            \\
Date       & Obs. ID/Epoch      	& $\Gamma$ 			& Norm. 			          & $\chi_{\nu}^{2}$ & Unabsorbed Flux  \\ 
       &       	&  			& 			          &  & ergs cm$^{-2}$ s$^{-1}$  \\ 
\tableline
2000-1-18  & Chandra 319 (bright) 	& 2.42$\pm^{0.7}_{0.6}$ 	& 5.81$\pm^{2}_{2}$$\times$10$^{-6}$    & 0.43 & 2.8$\times$10$^{-14}$ \\
2000-1-18  & Chandra 319 (faint) 	& 2.93$\pm^{2}_{1}$ 	& 5.53$\pm^{4}_{4}$$\times$10$^{-7}$    & 0.31 & 2.8$\times$10$^{-15}$\\
2001-1-07  & XMM 0055140101 	        & 2.47$\pm^{0.9}_{0.6}$ 	& 8.91$\pm^{4}_{4}$$\times$10$^{-6}$    & 0.29 & 4.2$\times$10$^{-14}$  \\
2003-2-13  & Chandra  2942  	        & 1.33$\pm^{1}_{1}$ 	& 2.21$\pm^{0.9}_{1}$$\times$10$^{-6}$     & 0.60 & 1.9$\times$10$^{-14}$\\
\enddata
\end{deluxetable}

\clearpage
\phantom{a}

\begin{figure}
\begin{center}
\includegraphics[width=60mm]{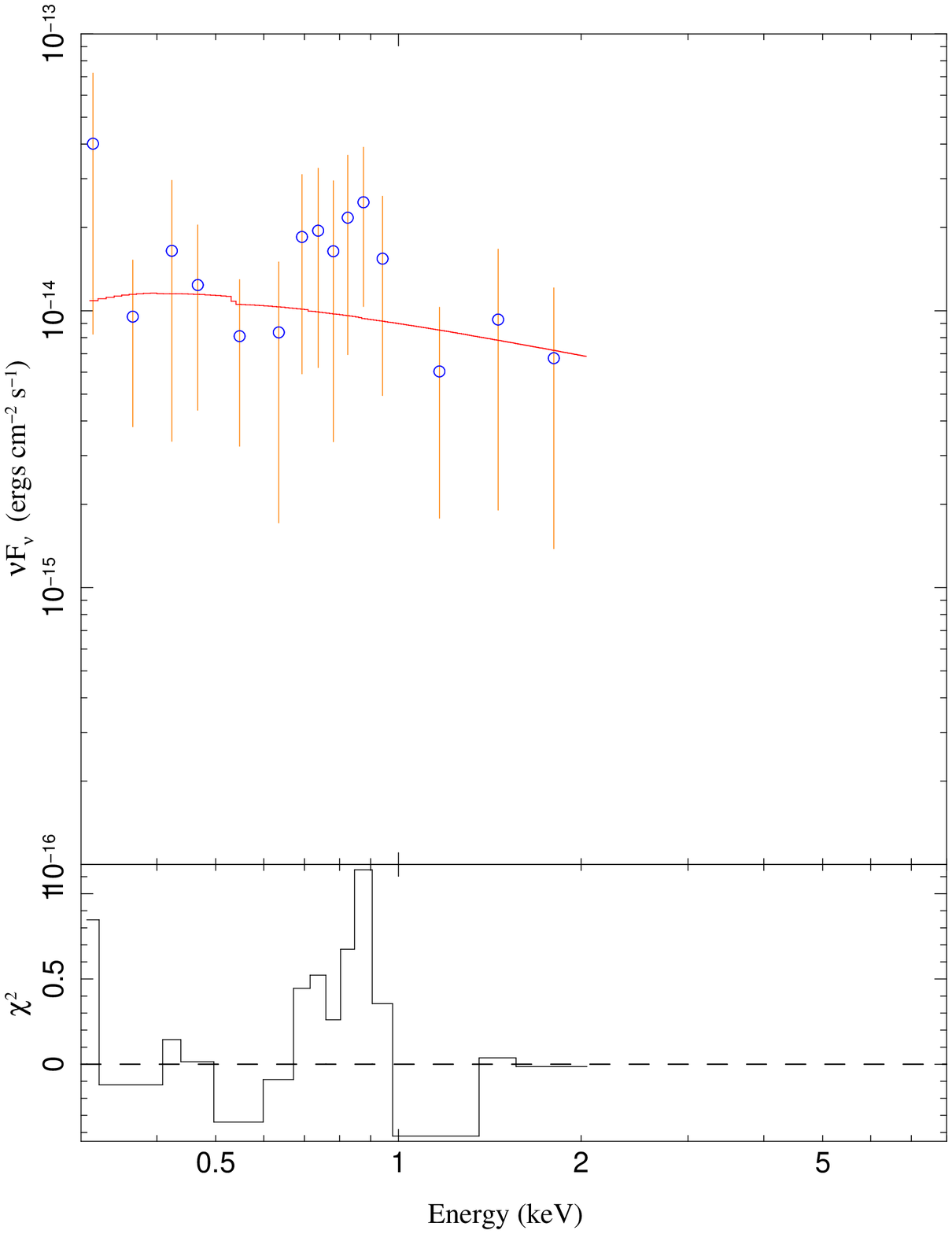}
\includegraphics[width=60mm]{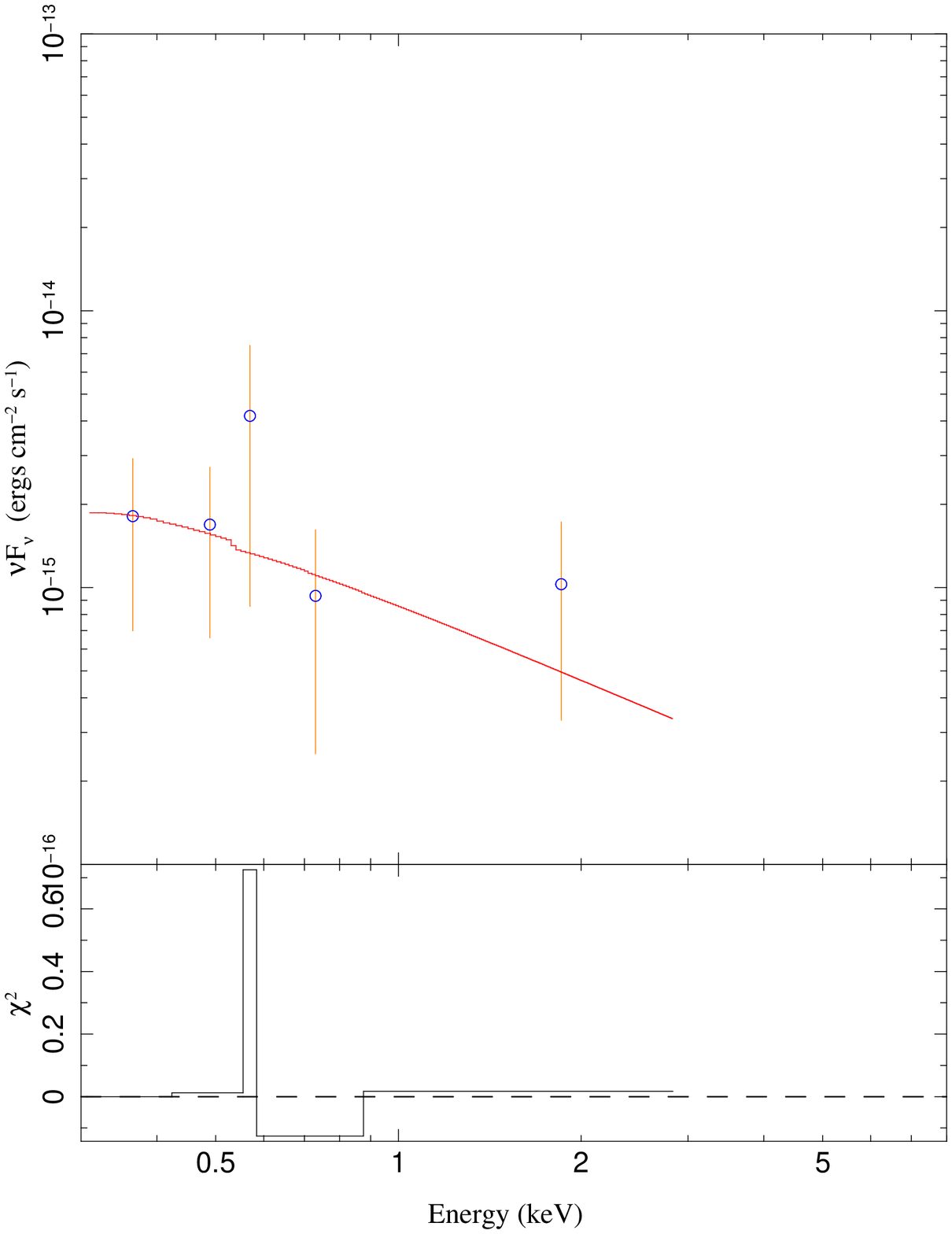}
\caption{The energy density distribution of CXOKMZJ033831.7-353058 during the bright (left) and faint (right) phase of the Chandra Obs ID 319 observations. The best fit powerlaw models and residuals are also shown. There is no obvious difference in the spectrum. }
\label{319 xid5 spectra}
\end{center}
\end{figure}

\end{document}